\documentstyle[psfig,aps,twocolumn,prb]{revtex}

\begin{document}
\wideabs{
\title{Superconducting gap parameters of MgB$_2$ obtained on MgB$_2$/Ag
and MgB$_2$/In junctions} \author{A.Plecenik and \v{S}.Be\v{n}a\v{c}ka}
\address{Institute of Electrical Engineering, Slovak Academy of Sciences,
          84239 Bratislava, Slovak Republic}
\author{P.K\'{u}\v{s} and M.Grajcar}
\address{Department of Solid State Physics, FMFI, Comenius University,
         84215 Bratislava, Slovak Republic }

\maketitle
\date{\today }

\begin{abstract}

MgB$_2$ superconducting wires with critical temperature T$_c$
approaching 40 K were used for preparation of MgB$_2$/Ag and
MgB$_2$/In junctions. The differential conductance $vs.$ voltage
characteristics of N-S junctions exhibit clear contribution of Andreev
reflection. Using modified BTK theory for s-wave superconductors two
order parameters $\Delta_{dirty}\approx 4$ meV and $\Delta_{3D}\approx
2.6$ meV have been determined from temperature
dependencies. Surprisingly, larger order parameter $\Delta_{dirty}$
vanishes at lower temperature $T_{c dirty}\approx 20$ K than smaller
one $\Delta_{3D}$ with $T_c\approx 38$ K. Both the magnitudes of the order
parameters and their critical temperatures  are in
good agreement with theoretical calculations of electron-phonon
coupling in MgB$_2$ carried out by Liu et al.
\end{abstract}
 }


The recent discovery of superconductivity in magnesium diboride
\cite{Nagamatsu}, with critical temperature T$_c$ about 40K, has
raised enormous interest.  There appear speculations about potential
applications, motivated by the simplicity of the crystal structure of
MgB$_2$ in comparison with the multicomponent high-T$_c$ copper-oxide
superconductors. A question arises what are the surface or interface
properties of this superconductor in contact with a normal metal, a
question extremely important for superconductor electronics.  Based on
this reason a lot of efforts is made to explain their physical
properties.  However, normal as well as superconducting parameters of
MgB$_2$, such as energy gap $\Delta$, Fermi velocity $v_F$ or
coherence length $\xi$ has not been exactly determined up to now.
Sharoni et al. \cite{Sharoni} measured the value of gap parameters in
the range of 5 to 7 meV and suggested that MgB$_2$ is a conventional
BCS $s$-wave superconductor. Karapetrov et al. \cite{Karapetrov}
measured the value of the gap about 5 meV and showed that temperature
dependence of the gap follows the BCS form.  However, Schmidt et
al. \cite{Schmidt} and Kohen and Deutcher \cite{Kohen} measured value
of the gap parameter about 4 meV and Rubbio-Bollinger et
al. \cite{Rubio-Bollinger} about 2 meV only.  In all cases, no evidence
for the energy gap anisotropy in the tunneling spectra has been
observed.

In this letter transport properties of MgB$_2$/In and MgB$_2$/Ag
contacts are discussed with the aim to contribute to the above
mentioned open questions.  It is shown that the temperature dependence
of the gap parameter with value of 4 meV do not follows BCS form but
rather coincides with two-gap model suggested by Liu et
al.\cite{Liu}

\begin{figure}[t]
\centerline{ \psfig{file=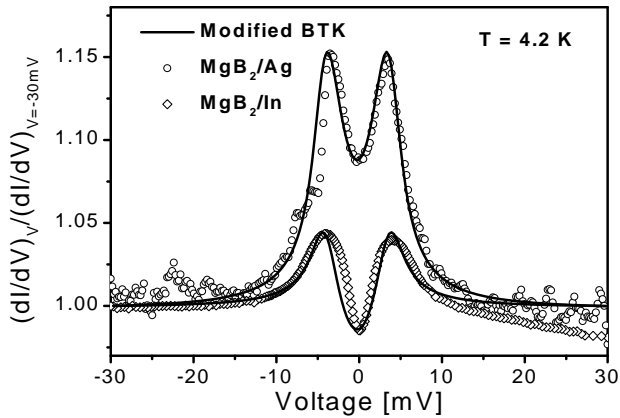,width=\columnwidth,angle=0} }
\caption{ Differential characteristics dI/dV vs. V measured on
MgB$_2$/Ag and MgB$_2$/In junctions at temperature T = 4.2 K
(symbols - experiments, solid lines - theoretical dependence
calculated by modified BTK theory). } \label{fig:GV}
\end{figure}

MgB$_2$ superconducting wires were produced by sealing of
150$\mu$m in diameter straight boron fiber with tungsten core
(approximately 15 $\mu$m diameter) and magnesium chips (99.8\% pure)
into a niobium tube with a component ratio Mg:B=3:1. The sealed Nb
tube was inserted into a quartz tube evacuated to 10 Pa. In order
to decrease the oxygen concentration in the vacuum background the
quartz tube was splashed many times with the argon (99.995\% pure)
before the final filling up with argon (0.3 atm.). Then it was
heated slowly up to 970$^o$C with a temperature steps ~8$^o$C/min and
kept at this temperature for two hours. The total time of
annealing process was four hours. After annealing the quartz tube
was removed from the furnace and quenched to room temperature in
approximately five minutes. The Nb tube was then removed from
the quartz tube and opened to air. MgB$_2$ wires, originally
elastic and straight, were after the above described procedure very
brittle.
\begin{figure}
\centerline{ \psfig{file=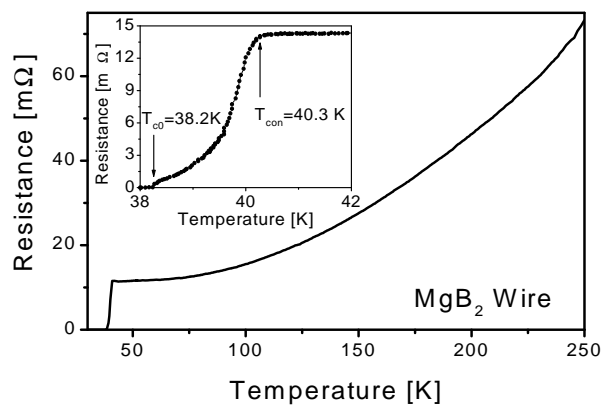,width=\columnwidth,angle=0} }
\caption{ R-T dependence of the MgB$_2$ wire. } \label{fig:RT}
\end{figure}
The junctions using MgB$_2$ wire were prepared by two methods. To
limit fragility, the MgB$_2$ wire was fixed at four points on the
Al$_2$O$_3$ substrate by Ag paste. These fixing points were used as
contacts for four point measurement of the resistance vs.  temperature
(R-T). Among them the Ag drops were located for completion of
MgB$_2$/Ag junctions. The differential characteristics were measured
between an Ag drop and the MgB$_2$ wire. The contacts were prepared
and measured immediately after the MgB$_2$ wire
preparation. MgB$_2$/In junctions were realized by traversing of a
freshly cleaved In wire across the MgB$_2$ wire (MgB$_2$ wires were
kept one day in air atmosphere at room temperature). The resistance of
the MgB$_2$ wire was measured by conventional four point method. The
differential characteristics dV/dI vs. V were measured by the
low-frequency (800 Hz) phase-sensitive detection technique, using a
resistance bridge. The dI/dV vs. V characteristics were obtained by
numerical inversion of the measured data.
The R-T dependence measured on MgB$_2$ wire is shown in Fig.~\ref{fig:RT}. This
dependence indicates a small inhomogeneity of the wire because
the width of the phase transition to superconducting state,
$\Delta$T$_c$, was about 2 K (see the inset in Fig.~\ref{fig:RT}).
Typical differential characteristics obtained on MgB$_2$/Ag and
MgB$_2$/In junctions are shown in Fig.~\ref{fig:GV} (symbols). These
characteristics are typical for N/S junctions (N is normal metal and S
is superconductor) with very high transparency of the tunnel
barrier.
Because the contribution of the Andreev reflection to the
total junction conductivity is small we suppose the existence of
normal metal regions on the MgB$_2$ surface. Thus the existence of
several conducting channels in the junction is possible. But in all
cases the superconducting regions are without degraded surface, as
indicated by the sharp differential characteristics. Both types of the
junctions, shown in Fig.~\ref{fig:GV}, may be well described by the
BTK model \cite{BTK}, where the total current is expressed as:

\begin{displaymath}
I_{NS}\sim\int_{-\infty}^{\infty}[f(E-eV)-f(E)][1+A(E)-B(E)]dE ,
\end{displaymath}
where $f(E)$ is Fermi distribution function, $A(E)$ and $B(E)$ are
coefficients giving probability of Andreev and ordinary
reflection, respectively. Due to small discrepancies between the
experimental and theoretical curves, for better fitting of
experimental curves, we used the modified BTK theory (taking into
the finite lifetime of the quasiparticles $\Gamma$)
\cite{Plecenik94}. The symbols in the modified BTK theory are as
follows: $A=aa^*$, $B=bb^*$, $a^*$, $b^*$ are the complex
conjugate quantities, $a=u_0v_0/\gamma$,
$b=(u_0^2-v_0^2)(Z_{eff}^2+iZ_{eff})/\gamma$, $\gamma=u_0^2+
(u_0^2-v_0^2)Z_{eff}^2$, $u_0$ and $v_0$ are the Bogoliubov
factors given as
$u_0^2=(1-v_0^2)=\frac{1}{2}\left[1+\sqrt{(E+i\Gamma)^2-\Delta^2}/
(E+i\Gamma)\right]$, and $Z_{eff}$ is the strength of the barrier.
We have fitted the differential characteristics measured on
MgB$_2$/Ag junctions because they were prepared immediately after
MgB$_2$ preparation and we expect small degradation due to the
influence of the atmosphere. From fitting of the experimental
curves (solid lines in Fig.~\ref{fig:GV}) the maximal value of the
gap parameter $\Delta$ = 4.2 meV, $Z_{eff}$ = 0.42 and $\Gamma$ =
0.8 meV were obtained. From the relation \cite{BT}
$Z_{eff}=\left(Z^2+(1-v_{F}/v_{N})^2/(4v_{F}/v_{N})\right)^{1/2}$,
where $v_{F}$ and $v_{N}$ are Fermi velocities of MgB$_2$ and
normal metal, respectively, the Fermi velocity of MgB$_2$ $v_{F} =
6.1\times10^7$ cms$^{-1}$ was calculated. From MgB$_2$/In
characteristics the Fermi velocity of MgB$_2$ $v_F =
4.7\times10^7$ cms$^{-1}$ was obtained.  Calculated Fermi
velocities of MgB$_2$ are in reasonable agreement with the band
structure calculation of Kortus et al. \cite{Kortus} Here we have
assumed that there is no artificial or native barrier (Z=0).
\begin{figure}
\centerline{ \psfig{file=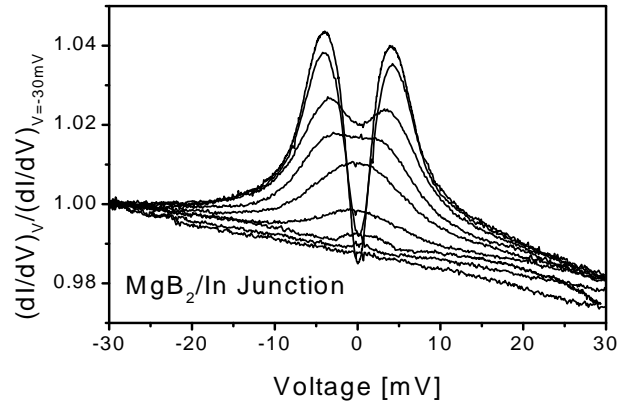,width=\columnwidth,angle=0} }
\caption{ Temperature dependence of differential characteristics
dI/dV vs. V measured on MgB$_2$/In junctions at temperatures T =
4.2, 8, 13, 18, 24, 28, 34, 37 and 40  K. } \label{fig:GT}
\end{figure}
Below T$_c$ the low - voltage differential conductance increases
with decreasing of temperature (Fig.~\ref{fig:GT}).  These curves
can be fitted by modified BTK theory.  The temperature dependence
of the gap parameter (taking into acount one energy gap) is shown
in Fig.~\ref{fig:DT} (open circles). Comparison of the gap
parameter obtained from experimental curves with the BCS
dependence (solid line 1 in Fig.~\ref{fig:DT}) shows large
discrepancy. In addition, to fit experimental curves one should
change the fitting parameters $Z$ and $\Gamma$ with temperature.
Also the ratio 2$\Delta(0)$/k$_B$T$_c$ is suprisingly small. For
$\Delta(0)$ = 4 meV and T$_c$ = 38 K (from MgB$_2$/In junction) we
obtained 2$\Delta(0)$/k$_B$T$_c$ = 2.4. A similar value was
obtained by tunneling spectroscopy by Kohen and Deutcher in Ref.
\onlinecite{Kohen}. These authors ascribed a small value of
2$\Delta(0)$/k$_B$T$_c$ to the existence of superconducting phase
with reduced critical temperature T$_c$ at MgB$_2$/Au interface.
However, from our spectra it is clear (see Fig.~\ref{fig:GT} and
Fig.~\ref{fig:DT}) that Andreev reflection disappers at the
critical temperature T$_c\approx 38$ K of the bulk material. On
the other hand there is deviation of the temperature dependence of
the energy gap from the BCS behavior at temperature $T\approx 20$
K. From the temperature dependence of energy gap determined within
one-gap model (Fig.~\ref{fig:DT}) one can see two characteristic
values of energy gap ($\approx$ 4 meV and $\approx$ 2.6 meV) and
two temperatures ($\approx$ 20 K and $\approx$ 38 K).
\begin{figure}
\centerline{ \psfig{file=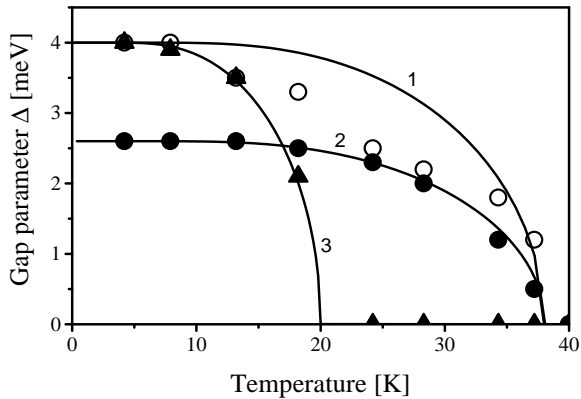,width=\columnwidth,angle=0} }
\caption{ Temperature dependence of gap parameters in one gap
approximation (open circles) and within multigap scenario in dirty
(solid triangles) and clean limits (solid circles) obtained from
experimental curves shown in Fig.~\ref{fig:GT}. The solid lines
1,2 and 3 corresponds to BCS relation $\Delta (T)/\Delta (0)=\tanh
(\Delta(T)T_c/\Delta(0)T)$} \label{fig:DT}
\end{figure}
These values correspond to those calculated by Liu et
al.\cite{Liu} within two-gap scenario. They argue that there are
two different order parameters in MgB$_2$ corresponding to two
sets of bands. One set of bands is quasi-2D with larger 2D gap
while 3D band exhibits 3D gap which is three times smaller in
magnitude. From a geometry of our tunnel junctions a tunneling in
direction perpendicular to B layers is expected i.e. the tunneling
spectra reflect 3D gap namely. On the other hand our samples are
not ideal. The large value of smearing parameter $\Gamma$
indicates dirty regions in contact area.  An enhanced defect
scattering in dirty regions leads to an averaging of both order
parameters that results to an isotropic order parameter and
reduced $T_c$.\cite{Liu} Thus our junction can be considered as a
parallel connection of two junctions corresponding to the
tunneling into the dirty and clean regions respectively. We have
fitted the differential characteristics according to above model
and we have obtained a good agreement between experimental and
theoretical curves.  The fitting procedure gives for dirty limit
$\Delta_{dirty} \approx$ 4 meV, $T_{c dirty} \approx$ 20 K and for
clean limit $\Delta_{3D}\approx$ 4 meV, $T_c \approx$ 38 K. Within
this model the fitting parameters $Z$ and $\Gamma$ were constant
in the whole temperature range. Temperature dependences of order
parameters $\Delta_{dirty}$ and $\Delta_{3D}$ are given in
Fig.~\ref{fig:DT}. Above values of the order parameters coinside
with those calculated in Ref.~\onlinecite{Liu}. In order to
measure also larger 2D order parameter, tunnel junctions should be
prepared in direction parallel to B layers. Such junctions cannot
be prepared by technology presented here but can be perhaps formed
on MgB$_2$ epitaxial thin films. The work is in progress.
Nevertheless there are already experimental results that give some
evidence on existance of the large value of order parameter. Very
recently Sharoni et al. \cite{Sharoni} measured the maximal value
of order parameter $\approx 7$~meV  by STM/STS which can be
ascribed to $\Delta_{2D}$.

In conclusion, we have presented properties of the MgB$_2$/Ag and
MgB$_2$/In junctions and an analysis of the measured differential
characteristics.  Our experimental data give an evidence on multigap
scenario in MgB$_2$.  From the fitting procedure the values of
$\Delta_{dirty}\approx 4$~meV, $T_{c dirty}\approx 20$~K and
$\Delta_{3D}\approx 2.6$~meV, $T_c\approx 38$~K, as well as an upper
limit of the Fermi velocity v$_F$ = 6.1$\times$10$^7$~cm/s were
determined.

\section*{Acknowledgement}
We would like to thank R. Hlubina for stimulating discussions,
 P. \v{S}ebo for help with a boron wire, and namely  I. I. Mazin for
calling our attention to Ref.~\onlinecite{Liu}. This work was
supported by the Slovak Grant Agency for Science (Grants No.2/7199/20
and 1/7072/20).


\begin{references}
\bibitem{Nagamatsu}J. Nagamatsu {\it et al.}, Nature {\bf 410}, 63 (2001).
\bibitem{Sharoni}A. Sharoni, I. Felner and O. Millo, cond-mat/0102325 (2001).
\bibitem{Karapetrov}G. Karapetrov {\it et al.}, cond-mat/0102312 (2001).
\bibitem{Schmidt}H. Schmidt {\it et al.},  cond-mat/0102389 (2001).
\bibitem{Kohen}A. Kohen and G. Deutscher, cond-mat/0103512 (2001).
\bibitem{Rubio-Bollinger} G. Rubio-Bollinger, H. Suderow and
S. Vieira, cond-mat/0102242 (2001).
\bibitem{Liu}Amy Y. Liu, I. I. Mazin and J. Kortus, cond-mat/0103570 (2001).
\bibitem{BTK}G. E. Blonder, M. Tinkham and T. M. Klapwijk, Phys.Rev.B
{\bf 25}, 4515 (1982).
\bibitem{Plecenik94}A. Plecenik {\it et al.}, Phys.Rev.B {\bf 49}, 10016 (1994).
\bibitem{BT}G. E. Blonder and M. Tinkham, Phys.Rev. B {\bf 27}, 112 (1983).
\bibitem{Kortus}J. Kortus {\it et al.},
cond-mat/0101446 (2001).

\end{references}
\end{document}